\documentclass[aps,prd,twocolumn,a4paper,superscriptaddress,nofootinbib]{revtex4-1}
\usepackage{graphicx}
\usepackage{multirow}
\usepackage{latexsym}
\usepackage[british]{babel}
\usepackage{soul}
\usepackage[colorlinks,linkcolor=blue,citecolor=blue,urlcolor=blue ]{hyperref}
\begin{document}

\newcommand{\be}{\begin{equation}}
\newcommand{\ee}{\end{equation}}
\newcommand{\bq}{\begin{eqnarray}}
\newcommand{\eq}{\end{eqnarray}}
\newcommand{\bsq}{\begin{subequations}}
\newcommand{\esq}{\end{subequations}}
\newcommand{\bc}{\begin{center}}
\newcommand{\ec}{\end{center}}

\title{Real-time cosmography with redshift derivatives}

\author{C. J. A. P. Martins}
\email{Carlos.Martins@astro.up.pt}
\affiliation{Centro de Astrof\'{\i}sica da Universidade do Porto, Rua das Estrelas, 4150-762 Porto, Portugal}
\affiliation{Instituto de Astrof\'{\i}sica e Ci\^encias do Espa\c co, CAUP, Rua das Estrelas, 4150-762 Porto, Portugal}
\author{M. Martinelli}
\email[]{martinelli@lorentz.leidenuniv.nl}
\affiliation{Institute Lorentz, Leiden University, PO Box 9506, Leiden 2300 RA, The Netherlands}
\affiliation{Institut f\"ur Theoretische Physik, Ruprecht-Karls-Universit\"at Heidelberg, Philosophenweg 16, 69120 Heidelberg, Germany}
\author{E. Calabrese}
\email{erminia.calabrese@astro.ox.ac.uk}
\affiliation{Sub-department of Astrophysics, University of Oxford, Keble Road, Oxford OX1 3RH, UK}
\author{M. P. L. P. Ramos}
\email[]{up201200278@fc.up.pt}
\affiliation{Centro de Astrof\'{\i}sica da Universidade do Porto, Rua das Estrelas, 4150-762 Porto, Portugal}
\affiliation{Faculdade de Ci\^encias, Universidade do Porto, Rua do Campo Alegre 687, 4169-007 Porto, Portugal}

\date{22 June 2016}

\begin{abstract}
The drift in the redshift of objects passively following the cosmological expansion has long been recognized as a key model-independent probe of cosmology. Here, we study the cosmological relevance of measurements of time or redshift derivatives of this drift, arguing that the combination of first and second redshift derivatives is a powerful test of the $\Lambda$CDM cosmological model. In particular, the latter can be obtained numerically from a set of measurements of the drift at different redshifts. We show that, in the low-redshift limit, a measurement of the derivative of the drift can provide a constraint on the jerk parameter, which is $j=1$ for flat $\Lambda$CDM, while generically $j\neq1$ for other models. We emphasize that such a measurement is well within the reach of the ELT-HIRES and SKA Phase 2 array surveys.
\end{abstract}
\maketitle

\section{Introduction}

One of the key paradigms established in cosmology in recent years is that the universe is seemingly dominated by two `dark' components, dubbed dark matter and dark energy, responsible for structure formation and for the recent accelerated expansion, respectively. Thus significant observational efforts are currently being put into characterizing this dark side of the universe. An important part of this endeavor consists in identifying specific tests discriminating between the simplest available model --the $\Lambda$ Cold Dark Matter model ($\Lambda$CDM)-- and its many alternatives \cite{Copeland,Clifton,Joyce}.

The drift in the redshift of objects which passively follow the cosmological expansion (called the redshift drift hereafter) has long been recognized as one of such tests \cite{1962ApJ136319S}. It is a direct non-geometric probe of the dynamics of the universe,
which doesn't rely on assumptions on gravity and clustering (other
than homogeneity and isotropy, if combining data from different lines of sight).
While currently available cosmological probes map our (present-day) past light cone, the redshift drift is unique in directly mapping the evolution by comparing past light cones at different times. As such, it can be used to distinguish between cosmological models \cite{Corasaniti:2007bg,Quercellini:2010zr}, and indeed it explores directions in parameter space that are often inaccessible and/or orthogonal to other observables, leading to degeneracy breaking and significantly improved constraints on cosmological parameters \cite{Martinelli:2012vq}.
 
Observational feasibility studies for redshift drift measurements have focused on forthcoming high-resolution ultra-stable optical/UV spectrographs~\cite{Loeb:1998bu,Liske:2008ph}, in particular with ELT-HIRES~\cite{HIRES}. These will enable measurements in the approximate redshift range $2<z<5$, thus
deep in the matter era. Observations at $z<1$ will likely be added by the SKA~\cite{Klockner:2015rqa} or 21cm experiments such as CHIME~\cite{Yu:2013bia}. Crucially, measurements with these different facilities rely on entirely different techniques (hence are vulnerable to different systematics) and they complement each other in redshift coverage. Therefore, for the first time we have the possibility of directly mapping the expansion history of the universe in the redshift range $0<z<5$.

One of our goals is to emphasize that the redshift drift also offers an independent way to measure quantities such as the Hubble, deceleration and jerk parameters which are crucial for the so-called cosmographic approach to cosmology \cite{Visser:2004bf}. While the first two are well constrained using other probes \cite{Riess:2016jrr}, the jerk parameter is still undetermined. As recently discussed in \cite{Bochner}, this stems from the fact that at very low redshifts, where simple --yet generic-- parametrizations of the jerk are sufficiently accurate, the data have not yet reached the needed accuracy; on the contrary at higher redshifts, despite the quality of the observations, the results will be less robust because of the strong dependence on the assumed parametrization. Other attempts to measure the jerk parameter and discussions about the difficulty of having accurate theory and data in the same redshift interval are presented in~\cite{Rapetti,Zhai,Mukherjee,Dantas}.

Here we explore the cosmological relevance of measurements of the derivative of the redshift drift (i.e., the second time derivative), providing illustrations of its discriminating power among different theoretical models. In the low-redshift limit, a measurement of the second time derivative informs us on the jerk parameter, which is $j=1$ for flat $\Lambda$CDM and generically $j\neq1$ for other models, providing a key test of the $\Lambda$CDM paradigm. In this work we focus on the theoretical derivation of the second derivative of the redshift but we also emphasize that such a measurement is well within the reach of the SKA Phase 2 array as well as of ELT-HIRES (albeit, in this case, with less sensitivity). Specific forecasts of the cosmological impact of these measurements will be addressed in a subsequent publication.

The paper is organized as follows. In Sec.~\ref{sec:two} we derive analytical expressions for the first and second derivative of the redshift, and in particular their low-redshift approximations. In Sec.~\ref{sec:three} we report examples for extended cosmological models and highlight some simple tests to confirm or rule out $\Lambda$CDM. We explore the sensitivity of future surveys in measuring redshift derivatives in Sec.~\ref{sec:four} and discuss the potential of these quantities in the concluding Sec.~\ref{sec:five}.

\section{Redshift derivatives}\label{sec:two}

Technological improvements will soon enable the observation of how the redshift of distant sources changes with time. In this section we derive the relevant equations for the first time derivative of the redshift, which have been presented in detail in \cite{Liske:2008ph}. We then generalize to a model independent framework the calculation of the second time derivative of the redshift (see \cite{1981ApJ24717L,Lake} for calculations in a specific model). We look specifically at the low-redshift limit of these quantities, which is appropriate for cosmography. Later, in Sec.~\ref{numerics} we will study the observationally relevant case of the redshift derivative of the drift. 

\subsection{First and second time derivatives}

We define the cosmological redshift, $z$, between the time of emission (represented by the scale factor of the emitter $a_{em}$) and the time of observation (represented by the scale factor of the observer $a_{obs}$) of a given signal as
\be
1+z=\frac{a_{obs}}{a_{em}}\,.
\ee
The derivative with respect to the time of observation can be written as 
\be
\frac{dz}{dt_{obs}}=\frac{1}{a_{em}}\frac{da_{obs}}{dt_{obs}}-\frac{1}{a_{em}}\frac{da_{em}}{dt_{em}}\,, \label{raw1st}
\ee
and, with the definition of the Hubble parameter
\be
H(z)=\frac{\dot a}{a}=H_{obs} E(z)\,,
\ee
(where the dot represents the derivative with respect to physical time) can be re-expressed as
\be\label{eq4}
\frac{dz}{dt_{obs}}=(1+z)H_{obs}-H(z) \,.
\ee
We can now write the dimensionless derivative of the redshift for photons observed at the present time ($t_{obs}=t_0$) and emitted at a generic earlier epoch ($t_{em}$=t), and get
\be\label{eq:z1}
Z_1(t_0,z)=\frac{1}{H_0}\frac{dz}{dt_0}=1+z-E(z)\,.
\ee

This calculation can easily be extended to compute the second time derivative. We start by differentiating Eq.~(\ref{raw1st}) and write
\bq
\frac{d^2z}{dt^2_{obs}}=\frac{1}{a_{em}}\frac{d^2a_{obs}}{dt^2_{obs}}-\frac{1}{a^2_{em}}\frac{da_{obs}}{dt_{obs}}\frac{da_{em}}{dt_{em}}\frac{dt_{em}}{dt_{obs}} + \nonumber\\
\frac{1}{a^2_{em}}\frac{da_{em}}{dt_{em}}\frac{dt_{em}}{dt_{obs}}\frac{da_{em}}{dt_{em}}-\frac{1}{a_{em}}\frac{d}{dt_{obs}}\left(\frac{da_{em}}{dt_{em}}\right)\,, \label{raw2nd}
\eq
which, with the additional definition of the deceleration parameter
\be
q(z)=-\frac{a{\ddot a}}{{\dot a}^2}=-\frac{\ddot a}{aH(a)^2}=-1+\frac{1}{2}(1+z)\frac{[E(z)^2]'}{E(z)^2}\,,
\ee
(where $'\equiv d/dz$) becomes
\be
\frac{d^2z}{dt^2_{obs}}=H_{obs}^2\left[\frac{1+q(z)}{1+z}E(z)^2-E(z)-q_{obs}(1+z)\right] \,.
\ee
Considering the present time and a generic earlier one, with the same definitions as above, the dimensionless second derivative is
\be
Z_2(t_0,z)=\frac{1}{H_0^2}\frac{d^2z}{dt_0^2}=\frac{1+q(z)}{1+z}E^2(z)-E(z)-q_0(1+z) \,,
\ee
or equivalently
\bq\label{eq:z2}
Z_2(t_0,z)&=&\frac{1+q(z)}{1+z}Z_1^2(z)-\left(1+2q(z)\right)Z_1\nonumber\\
&&+\left(q(z)-q_0\right)(1+z)\,.
\eq

An additional alternative (and somewhat more compact) expression can be given, using
\be
{\dot H(z)}=-[1+q(z)]H(z)^2 \label{eq9}
\ee
and
\be
H'(z)=\frac{1+q(z)}{1+z}H(z) \,,
\ee
leading to
\be 
Z_2(t_0,z)=(1+z)\left[E'-E'_0 \right]+Z_1(t_0,z)\left(1-E'\right)
\ee
or alternatively
\be
Z_2(t_0,z)=\frac{1}{2}(E^2)'-(1+z)E'_0+Z_1(t_0,z) \,,
\ee
where $E'_0$ is the derivative of $E$ computed at redshift zero.

\subsection{Low-redshift limit: real-time cosmography}

Assuming that many future redshift surveys will provide a large amount of low-redshift high-sensitivity data, it is convenient to derive here generic expressions for the low-redshift limits of $Z_1$ and $Z_2$ (we drop from now on the explicit dependence on the redshift). Taking the low-redshift limit of Eq. \ref{eq4}
\be
\frac{dz}{dt_0}\approx H_0(1+z)-H_0\left(1+\frac{H_0'}{H_0}z\right)=H_0z+\frac{\dot H_0}{H_0}z\,,
\ee
together with Eq.~(\ref{eq9}) and the fact that for a generic quantity $x$
\be
x'=-\frac{\dot x}{H(z)(1+z)} \,,
\ee
we obtain
\be
{\dot z}=-H_0q_0z \,,
\ee
or simply
\be\label{eq:lowz1}
Z_1=-q_0z \,.
\ee
This can be extended to $Z_2$ using the definition (discussed for example in \cite{Visser:2003vq}) of the jerk parameter
\be
j(a)=\frac{\dddot a}{aH(a)^3}=\frac{{\dddot a}a^2}{{\dot a}^3} \,,
\ee
which is the next term in the sequence
\be
{\dot a}=Ha\,,\quad {\ddot a}=-qH^2a\,,\quad {\dddot a}=jH^3a\,.
\ee
The jerk can also be written in the computationally useful form
\bq
j(z)&=&(1+z)^2\frac{E(z)''}{E(z)}+q(z)^2 \\
&=&\frac{1}{2}(1+z)^2\frac{[E(z)^2]''}{E(z)^2}-(1+z)\frac{[E(z)^2]'}{E(z)^2}+1\,.
\eq
We can now use the above expression to derive 
\be
{\dot q}=-(j-q-2q^2)H \,,
\ee
which, in the low-redshift limit, yields 
\be
{\ddot z}=j_0H_0^2z \,,
\ee
or more simply
\be\label{eq:lowz2}
Z_2=j_0z\,.
\ee
Therefore, in this low-redshift approximation
\be
Z_2\mp Z_1=(j_0\pm q_0)z\,.
\ee

We see that measurements of $Z_1$ and $Z_2$ directly yield, at least in principle, the present-day values of the deceleration and jerk parameters which are necessary for a fully model independent cosmographic approach to cosmology. In particular, as we will confirm in the following section, the value of the jerk provides a discriminating test between flat $\Lambda$CDM and other models. The practical details of such a measurement will be discussed in Sec.~\ref{sec:four}.

\section{A worked example: discriminating between $\Lambda$CDM and simple extensions}\label{sec:three}

It is instructive to discuss examples of specific models. These will provide illustrations of the behavior of the cosmographic parameters in these cases, showing how they relate to the cosmological parameters and therefore how they contribute to constrain the cosmological model. It also serves to derive in a more systematic way some results that have been used in the more phenomenological cosmography approach \cite{pheno0,pheno1,pheno2,pheno3,pheno4,pheno5}.

We start with a single fluid with an equation of state $p=w\rho$, with $w$ being a constant. In this 
case we have
\be
E^2(z)=(1+z)^{3(1+w)},
\ee
and consequently
\be
q=\frac{1}{2}(1+3w),
\ee
\be
j=1+\frac{9}{2}w(1+w),
\ee
so both parameters are constant.

We now consider the $w$CDM cosmological model, where to the standard cold dark matter and curvature components we add a dark energy component fully characterized by its equation of state parameter $w$. For this model we have
\be
E^2(z)=\Omega_m(1+z)^3+\Omega_d(1+z)^{3(1+w)}+\Omega_k(1+z)^2 \,,
\ee
where $\Omega_m$ is the present matter density, $\Omega_d$ the dark energy density and $\Omega_k$ the curvature term.\\
In this case we can trivially compute
\be
q(z)=\frac{1}{2}\frac{\Omega_m(1+z)^3+(1+3w)\Omega_d(1+z)^{3(1+w)}}{\Omega_m(1+z)^3+\Omega_d(1+z)^{3(1+w)}+\Omega_k(1+z)^2} \,,
\ee
\be\label{jerkwcdm}
j(z)=1+ \frac{1}{2}\frac{9w(1+w)\Omega_d(1+z)^{3(1+w)}-2\Omega_k(1+z)^2}{\Omega_m(1+z)^3+\Omega_d(1+z)^{3(1+w)}+\Omega_k(1+z)^2}   \,;
\ee
while for the time derivatives of the redshift we find
\be
Z_1=1+z-E(z) \,,
\ee
and
\bq
Z_2&=&(1+z)\left[1+\frac{3}{2}\Omega_m z +\frac{3}{2}(1+w)\Omega_d\left((1+z)^{1+3w}-1\right)\right] \nonumber\\
&&-E(z) \,.
\eq

The low-redshift limits of these time derivatives, expressed now in terms of the cosmological parameters, are
\be
Z_1=-\frac{1}{2}\left(\Omega_m+(1+3w)\Omega_d\right)\, z +O(z^2) \,,
\ee
\be
Z_2=\left[1-\Omega_k+\frac{9}{2}\Omega_d w(1+w)\right]\, z+O(z^2) \,,
\ee
and it is straightforward to verify that they reduce to Eqs.~(\ref{eq:lowz1}) and (\ref{eq:lowz2}) previously derived. 

Note that these low-redshift limits depend on both the matter and the dark energy content of the universe (including the equation of state of the latter). This is not the case at high-redshift where (still assuming only matter plus dark energy plus curvature, thus ignoring radiation) the derivatives reduce to 
\be\label{eq:now1}
Z_1\longrightarrow -\sqrt{\Omega_m}\, z^{3/2} \,,
\ee
\be\label{eq:now2}
Z_2\longrightarrow \frac{3}{2} \Omega_m\, z^2\,.
\ee
Here, both quantities behavior extends deep in the matter era, and, as expected, depends only on the matter density. Measurements of redshift derivatives at intermediate-to-high redshifts thus provide additional constraining power by characterizing the matter component (and therefore isolating the dark energy one at low redshift). This highlights the importance of complementary redshift drift measurements deep in the matter era \cite{Liske:2008ph,Martinelli:2012vq,Vielzeuf:2012zd}, to be carried out by ELT-HIRES \cite{HIRES}.

We also note that in the case of a flat universe the expression for the jerk, Eq.(\ref{jerkwcdm}), simplifies to
\be
j(z)=1+\frac{9w(1+w)(1-\Omega_m)}{2\left[1-\Omega_m\left(1-(1+z)^{-3w}\right)\right]}\,,
\ee
from which we observe that, for flat $\Lambda$CDM, $j=1$ holds for {\it every} redshift. Thus a direct measurement of the jerk term (at any low redshift) is a powerful discriminating test between flat $\Lambda$CDM and its alternatives. This is of course not applicable to intermediate-to-high redshift because of the contribution from radiation.

Finally, we briefly consider an example of a dark energy with a dynamical equation of state, choosing the CPL parametrization \cite{CPL1,CPL2}
\be
w(z)=w_0+w_a\frac{z}{1+z}\,.
\ee
In this case $q_0$ can be written as
\be
q_0=\frac{1}{2}\left[\Omega_m+(1+3w_0)\Omega_d \right]
\ee
(note that this does not depend on $w_a$) and $j_0$ as
\be
j_0=1-\Omega_k+\frac{9}{2}\Omega_d w_0(1+w_0)+\frac{3}{2}\Omega_dw_a\,.
\ee

It is now interesting to notice that the sum of the two reads
\be
q_0+j_0=\frac{1}{2}(3\Omega_m-\Omega_d)+\frac{1}{2}(2+3w_0)^2\Omega_d+\frac{3}{2}w_a\Omega_d\,.
\ee
This implies that for the particular case of $\Lambda$CDM we have the interesting relation
\be
q_0+j_0=\frac{3}{2}\Omega_m \,,
\ee
regardless of the curvature $\Omega_k$. Once again, a measurement of these terms appears to be a powerful consistency test for $\Lambda$CDM.

\section{\label{sec:four}Numerical estimate of future data sensitivity}

What observations of the redshift drift actually measure is the shift in the spectroscopic velocity of a source ($\Delta v$) in a given time interval ($\Delta t$). This shift is related to the first time derivative of the redshift via
\be
\Delta v = \frac{c\Delta z}{1+z}=cH_0\Delta t\frac{Z_1}{1+z}
\ee
where $c$ is the speed of light. 

Upcoming experiments such as the E-ELT and SKA will achieve, through different means, high enough spectroscopic sensitivity to measure this velocity shift. The E-ELT's high-resolution optical spectrograph will be able to measure the shift in the spectroscopic velocity, observing the Lyman $\alpha$ absorption lines of distant quasar systems, in a redshift range $2<z<5$ \cite{Loeb:1998bu,HIRES}, while SKA will measure $\Delta v$ through observations of the neutral hydrogen (HI) emission signal of galaxies at two different epochs to a precision of a percent (in redshift space) in the range $0<z<1$ (for the SKA Phase 2 array) \cite{Klockner:2015rqa}. Note that the two experiments ideally complement each other, with the E-ELT probing the deep matter era while the SKA probes the acceleration era and its onset.

\subsection{\label{numerics}Numerical redshift derivatives from data}

We extend previous works which explored measurements of the redshift drift and infer the sensitivity of these experiments to the second time derivative of redshift by resorting to the measurement of $Z_1$ at several different redshifts. We emphasize that a measurement of $Z_2$ provides additional information which is not contained in $Z_1$, though there is currently no feasible way of directly measuring $Z_2$. However, a quantity closely related to $Z_2$ can in fact be obtained through the numerical redshift derivative of $Z_1$
\be
\frac{dZ_1(t_0,z)}{dz} = 1 - E(z)'= -q(z)+Z_1(t_0,z) \frac{1+q(z)}{1+z} \,, \label{eq:z1der}
\ee
or equivalently
\be
\frac{dZ_1(t_0,z)}{dz} = 1+\frac{1+q(z)}{1+z} [Z_1-(1+z)]  \,.
\ee

This expression can be inverted, allowing us to express the deceleration parameter at any redshift as a function of $Z_1$ and $Z'_1=dZ_1/dz$
\be
q(z) = -1+\frac{1-Z'_1}{1+z-Z_1}  \,;
\ee
similarly, for the jerk parameter we have
\be
j(z) = q^2(z)-\frac{(1+z)^2Z''_1}{1+z-Z_1}  \,.
\ee

This is relevant because in the low-redshift (linearized) limit $Z_2$ and $dZ_1/dz$, although different, contain the same cosmographic information. Specifically, $dZ_1/dz$ has the form 
\be\label{eq:lowzD}
\frac{dZ_1(t_0,z)}{dz} \sim -q_0+(q_0^2-j_0)z+O(z^2)\,,
\ee
which again only depends on the cosmographic parameters $q_0$ and $j_0$. We note that this expression is fully generic (other than the assumption of a metric theory of gravity), so the constraints coming from its measurement would be fully model independent.

\subsection{\label{skaeltexamples}SKA and E-ELT scenarios}

Assuming to have $N$ measurements of $\Delta v$ at some redshifts $z_i$, one can obtain $N-1$ measurements of $dZ_1/dz$ as numerical derivatives
\be\label{eq:dz1dz}
D(\bar{z})\equiv\frac{dZ_1}{dz}(\bar{z}) = \frac{Z_1(z_{i+1})-Z_1(z_i)}{z_{i+1}-z_i}
\ee
with $\bar{z}=(z_{i+1}-z_i)/2$.\\
The errors on these measurements can be obtained as
\be\label{eq:Derr}
\frac{\sigma^2}{D^2}=\frac{\sigma^2_{Z_1(z_{i+1})}+\sigma^2_{Z_1(z_{i})}}{(Z_1(z_{i+1})-Z_1(z_i))^2} + \frac{\sigma^2_{z_{i+1}}+\sigma^2_{z_{i}}}{(z_{i+1}-z_i)^2}
\ee
where $\sigma_{Z_1(z_{i})}$ is obtained propagating the error on $\Delta v$ as
\be\label{eq:prop}
\sigma_{Z_1}=\sqrt{\left(\frac{\partial Z_1}{\partial \Delta v}\right)^2\sigma^2_{\Delta v}+\left(\frac{\partial Z_1}{\partial H_0}\right)^2\sigma^2_{H_0}+
                   \left(\frac{\partial Z_1}{\partial z}\right)^2\sigma^2_{z}} \,,
\ee
which we rewrite as 
\be
\sigma_{Z_1}=\sqrt{\Sigma^2_{\Delta v}+\Sigma^2_{H_0}+\Sigma^2_{z}} \,.
\ee
It is worth noticing that the uncertainty in the Hubble constant $H_0$ also contributes to the overall error budget; in this paper we use the uncertainty on $H_0$ obtained by Planck \cite{Ade:2015xua} measurements $\sigma_{H_0}=0.66$. Furthermore, we assume $\sigma_z=0.001$ for the SKA, while we assume a negligible redshift error for the high resolution spectrograph of the E-ELT \cite{HIRES}.

For the SKA we adopt the analysis and estimates discussed in \cite{Klockner:2015rqa}. These assume that the frequency shift in redshift space can be established to a 
precision of a percent. For a drift signal of order centimeters per second per year, this requires a precision of $10^{-3}$Hz, leveraging the SKA sensitivity and number counts. This leads us to consider the following two scenarios
\begin{enumerate}
\item 
For SKA Phase 1, three measurements of the drift $\Delta v$ in redshift bins centered on $z_i=[0.1,0.2,0.3]$ with velocity uncertainties $\sigma_v$ respectively of $3\%$ in the first 
bin, $5\%$ in the second and $10\%$ on the third. Achieving such an uncertainty in Phase 1 
will require a timespan of 40 years (although long this is within the expected full SKA timespan of 50 years).
Despite this long integration time, we will use this as a benchmark scenario to shed light on the gains brought by the improvements in sensitivity and redshift coverage afforded by the full SKA configuration.
\item For SKA Phase 2, we adopt a configuration with ten measurements of the drift $\Delta v$ in equally spaced redshift bins with centers from $z=0.1$ to $z=1.0$, and with velocity uncertainties $\sigma_v$ ranging from $1\%$ to $10\%$. This could be reached in only 0.5 years, leading to an extremely competitive, ideal, scenario. We however notice that to achieve this configuration $10^7$ galaxies are required in each bin. (For plotting purposes we will also consider an alternative with five equally spaced redshift bins with centers from $z=0.2$ to $z=1.0$, with the same timespan of 0.5 years. We assume that in this case the error on the measured velocity will be reduced by a factor $\sqrt{2}$ with respect to the ten bins configuration.)
\end{enumerate}

The E-ELT, as discussed in \cite{Liske:2008ph}, is expected to observe the shift in spectroscopic velocity with an uncertainty 
(in centimeters per second)
\be
\sigma_{\Delta v} = 1.35 \frac{2370}{S/N}\sqrt{\frac{30}{N_{QSO}}}\left(\frac{5}{1+z_{QSO}}\right)^x
\ee
with $x=1.7$ for $z\leq4$ and $x=0.9$ beyond that redshift. In what follows we will assume a signal-to-noise ratio $S/N\approx3000$, $N_{QSO}=10$ quasars for each of three redshift bins at redshifts 
$z_i=[2.5,3.5,5.0]$ and a timespan of $\Delta t\approx20$ years.

\begin{figure}[t!]
\begin{center}
\begin{tabular}{ccc}
\includegraphics[width=\columnwidth]{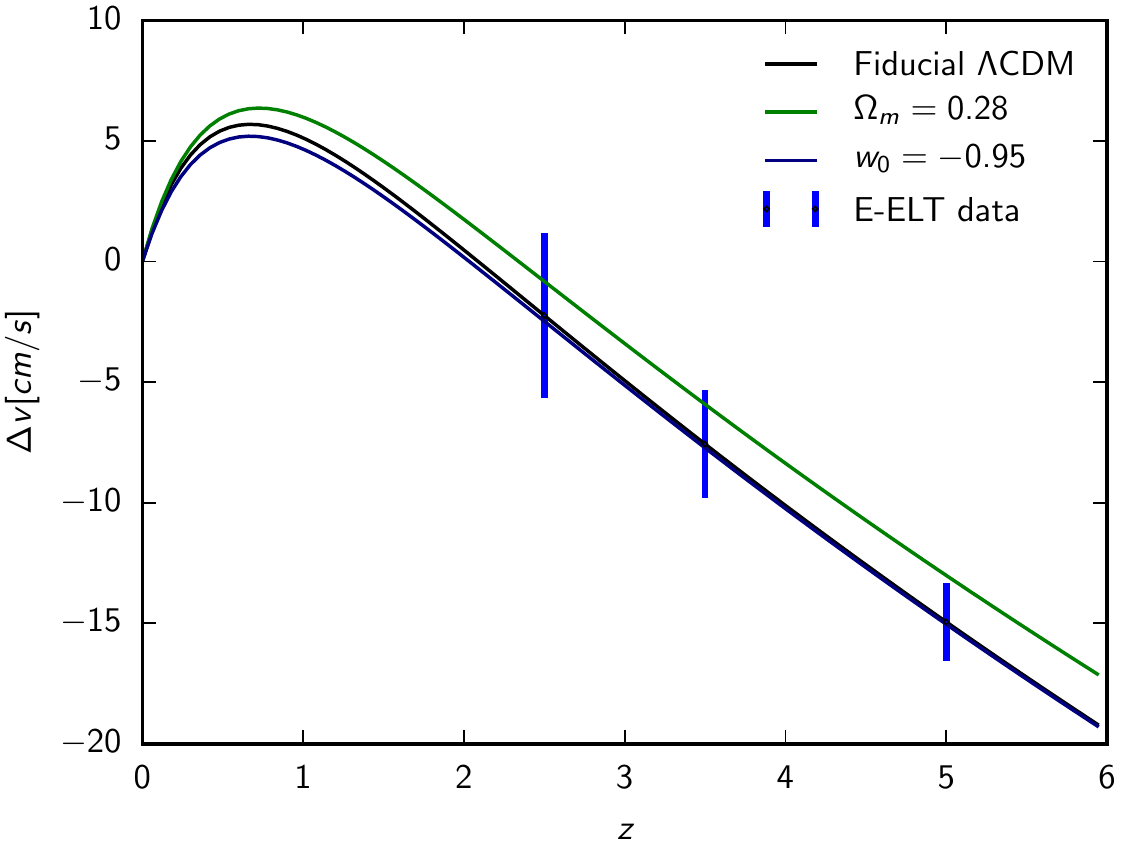}\\
\includegraphics[width=\columnwidth]{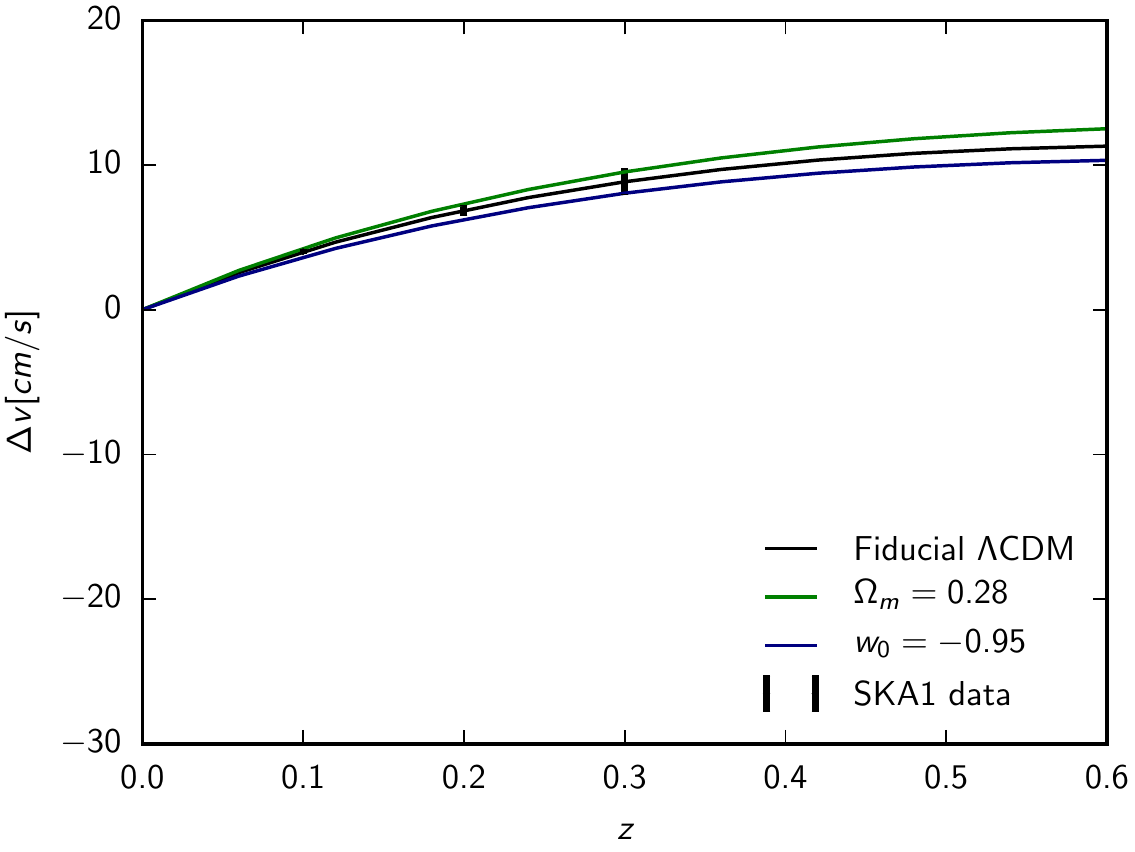}\\
\includegraphics[width=\columnwidth]{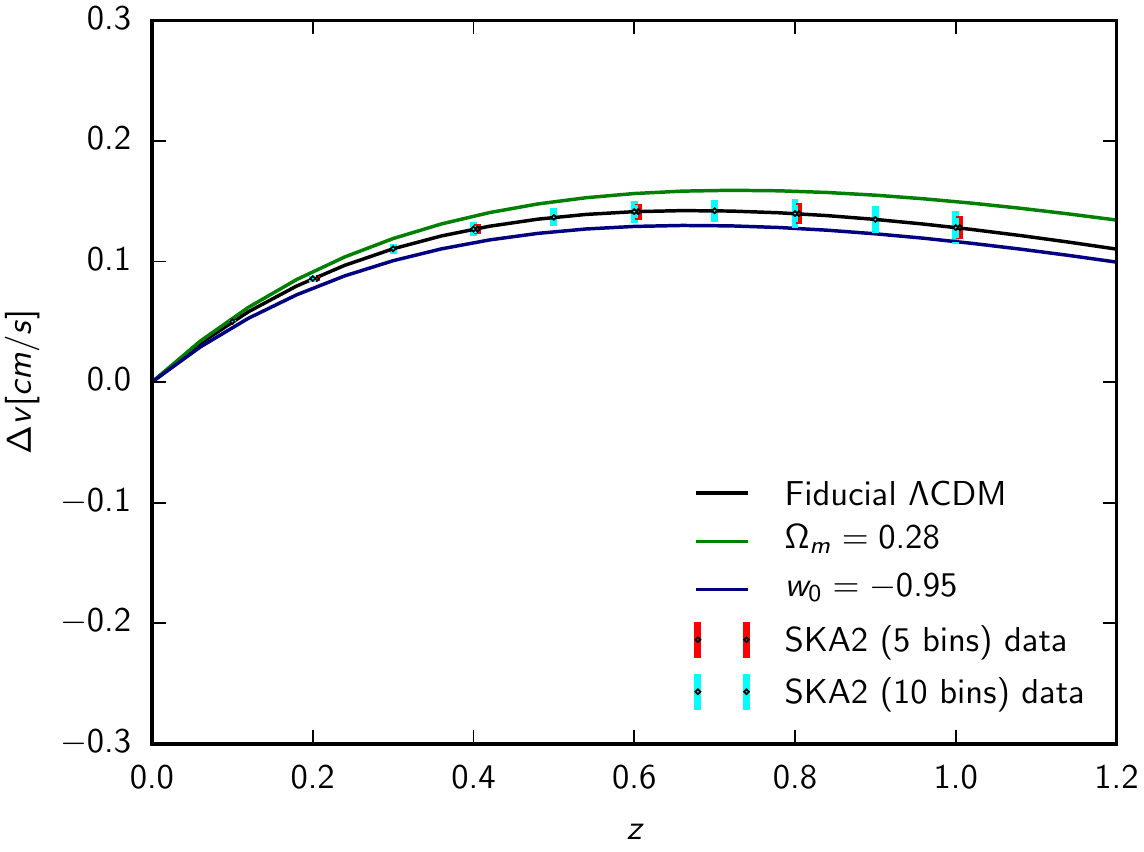}
\end{tabular}
\end{center}
\caption{$\Delta v$ measurements as expected from E-ELT (top panel), SKA1 (central panel) and SKA2 (bottom panel). The plots also show the theoretical $Z_1$ for a fiducial $\Lambda$CDM model with $w=-1$ and $\Omega_m=0.3$ (black line) and two alternative cosmologies, with $\Omega_m=0.25$ (green line) and with $w=-0.95$ (blue line).}\label{fig:deltav}
\end{figure}

\begin{figure}[ht!]
\begin{center}
\begin{tabular}{cc}
\includegraphics[width=\columnwidth]{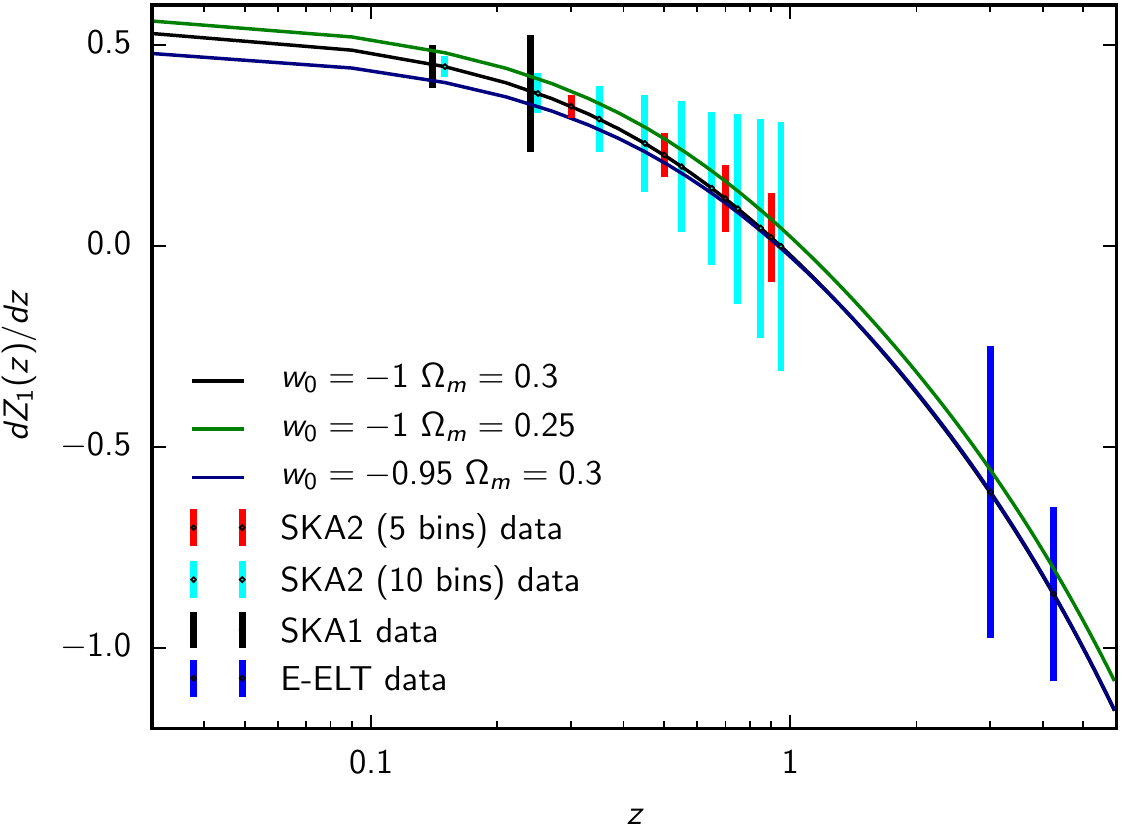}
\end{tabular}
\end{center}
\caption{Measurements of the $Z_1$ derivative as expected from E-ELT (blue error bars), SKA1 (black error bars) and SKA2 (red and cyan error bars). The plot also shows the theoretical $Z_1$ derivative for a fiducial $\Lambda$CDM model with $w=-1$ and $\Omega_m=0.3$ (black line) and two alternative cosmologies, with $\Omega_m=0.25$ (green line) and with $w=-0.95$ (blue line).}\label{fig:dZ1}
\end{figure}

\begin{figure}[ht!]
 \begin{center}
 \begin{tabular}{cc}
  \includegraphics[width=\columnwidth]{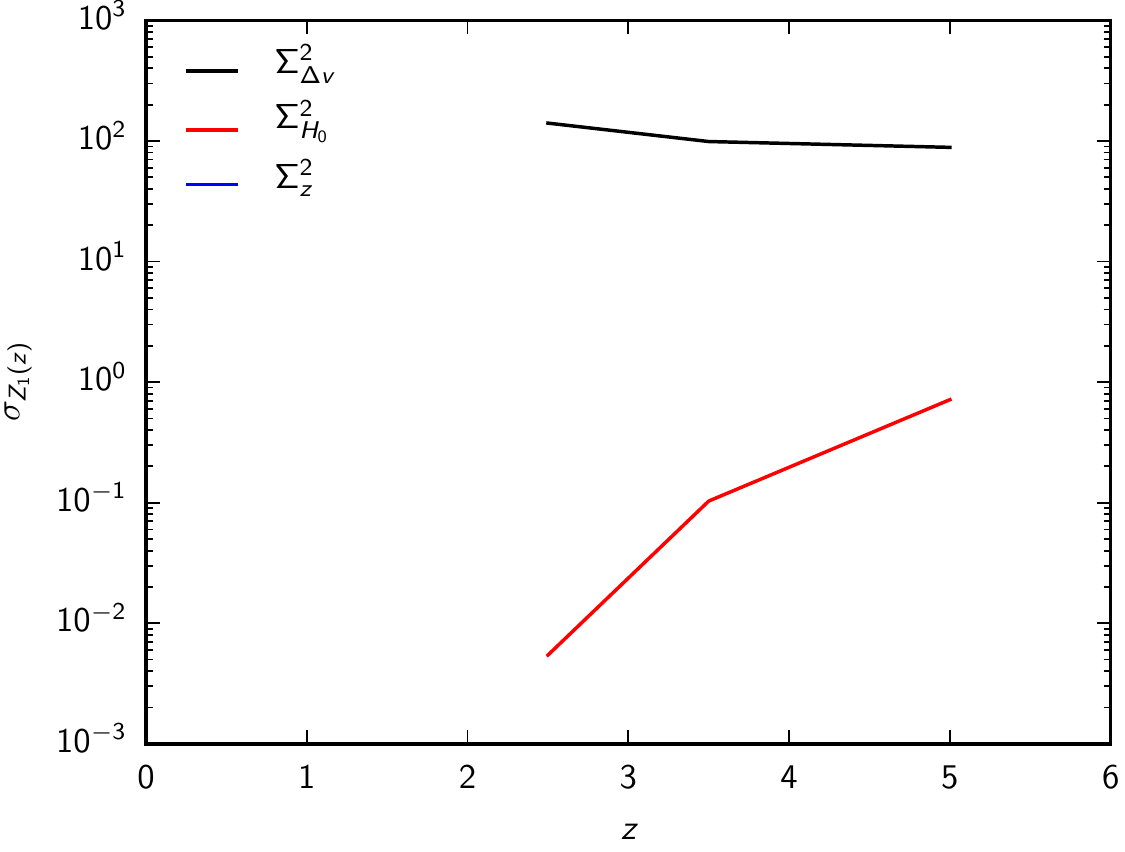}\\
  \includegraphics[width=\columnwidth]{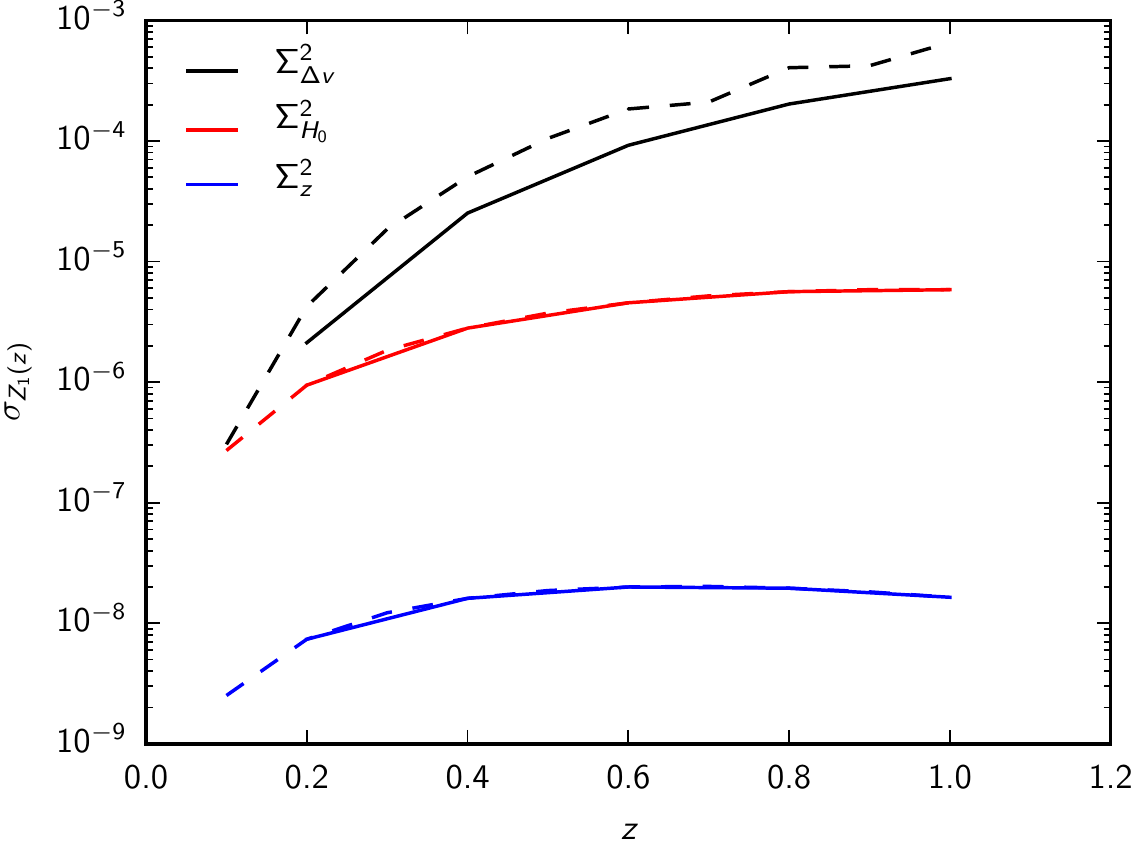}
 \end{tabular}
 \caption{Comparing the various $\Sigma$ contributions to the uncertainty in the E-ELT (top panel) and SKA2 (bottom panel) measurements of $Z_1(z)$. In the bottom panel solid lines refer to 5 bins SKA2 configuration while dashed lines refer to the 10 bins one.}\label{fig:noise}
 \end{center}
\end{figure}

Figure~\ref{fig:deltav} shows the $\Delta v$ (or equivalently the redshift drift) measurements expected from the two experiments; it is important to notice the big difference in the signal amplitudes, due to the different time interval $\Delta t$ reached by the two surveys.

Following the previous discussion and derivations, we now extend the data products of these experiments and include measurements of the second redshift derivative, $dZ_1/dz$. The expected errors on these measurements for E-ELT and SKA are shown in Figure \ref{fig:dZ1}. 

Comparing qualitatively the forecasted data with expectations from $\Lambda$CDM and different cosmologies, one sees that both the ELT and SKA can distinguish between models at high significance with $Z_1$ and SKA2 data will provide competitive $Z_2$ observations as well. However we want to stress that, because of the different dependence on the cosmological parameters, future tests of the cosmological model will benefit from the measurements of both quantities from both facilities.

Figures~\ref{fig:deltav}-\ref{fig:dZ1} also highlight how low redshift measurements are crucial to distinguish between different Dark Energy models, by noticing that the impact of the equation of state parameter $w$ vanishes at high redshift, as discussed in Eqs.~(\ref{eq:now1}-\ref{eq:now2}).

Finally, figure \ref{fig:noise} shows the various contributions to $\sigma_{Z_1}$. As expected, the error on $Z_1$, and therefore on $dZ_1/dz$, is dominated by the uncertainty in $\Delta v$ for E-ELT. When zooming into low redshifts instead, as in the case of SKA, also the contribution of $\sigma_{H_0}$ becomes significant.\\

As stated in Section~\ref{sec:two}, the low redshift limit of the drift can give direct information on the cosmography parameters $q_0$ and $j_0$. As a simple exercise, we can estimate the errors on these two parameters through error propagation from the forecasted data points of Figures~\ref{fig:deltav}-\ref{fig:dZ1} for SKA1 and SKA2 (10 bins case). In both cases we use only data with $z\leq0.3$ (which is the interval where the low redshift approximation is more suitable). In the former case we find
\be
\sigma_{q_0}\sim1.8\times10^{-2}\,,\qquad \sigma_{j_0}\sim 0.34\,, 
\ee
while in the latter these improve to
\be
\sigma_{q_0}\sim0.6\times10^{-2}\,,\qquad \sigma_{j_0}\sim 0.13\,. 
\ee
We emphasize that this is only a first approximated evaluation of the constraints on these parameters: to achieve a more robust estimation a more reliable analysis of SKA errors is needed, together with a detailed study of the validity of the low redshift approximation of Eqs.~(\ref{eq:lowz1}) and (\ref{eq:lowzD}).

\section{Conclusions}\label{sec:five}

The redshift drift of objects passively following the cosmological expansion has been shown to be a key model-independent probe of cosmology. At the conceptual level, it is the first probe to allow us to see the universe expanding in real time. At the practical level, it is independent from other experiments, and in particular its orthogonality to standard probes in cosmological parameter space helps breaking degeneracies between cosmological scenarios.

In this work we have presented new calculations of the first and second redshift derivatives and derived their simple approximations in the low-redshift limit. We have mapped these quantities into cosmographic and cosmological parameters and discussed how they provide a model-independent test of the expansion of the universe. In particular, the second redshift derivative contains information on the deceleration ($q_0$) and jerk ($j_0$) cosmographic parameters. We have pointed out that the latter parameter is a key discriminant between flat $\Lambda$CDM and alternative models.

Our main conclusion is that while $Z_1$ and $Z_2$ are the physically natural observables, being model independent (with the usual caveats) and measurable at any redshift (given a sufficiently stable detector and enough telescope time), in the particular case of low redshift measurements they directly yield the usual cosmography parameters $q_0$ and $j_0$. We also emphasize that $Z_2$ encodes information that is distinct from $Z_1$, but it is difficult to measure; on the other hand $dZ_1/dz$ is easier to measure, and (at low redshift) is closely related to $Z_2$. At least in principle, $Z_1$ and its first derivative $Z_1'$ allow a determination of the deceleration parameter $q(z)$ at any redshift, and similarly the addition of the second derivative $Z_1''$ allows the determination of the jerk, $j(z)$. In practice (in other words, observationally), the interesting question is until what redshift can such measurements be made such that the error coming from this low redshift series expansion is subdominant compared to the other statistical and systematic observational uncertainties: this is the scenario in which $q_0$ and $j_0$ can be directly measured. One can of course extend the redshift range (enabling the use of additional data), at the cost of including higher-order terms in the expansion, but in this case additional parameters will need to be fitted and it is not clear how that will impact the constraints on $q_0$ and $j_0$. Answering this question will require detailed realistic simulations of SKA data, and we note that the answer should also depend on the redshift dependence of the number density of galaxies.

We have demonstrated that the measurements of the redshift drift, $Z_1$, expected from the E-ELT and SKA surveys, also enable the numerical determination of its derivative, $dZ_1/dz$.  Our work highlights the complementary of the E-ELT and the SKA in mapping the expansion history of the universe in a model-independent way: using different observational techniques, the two experiments will probe different redshift ranges allowing for a direct reconstruction of the expansion of the universe both in the dark energy and matter dominated epochs. We have qualitatively demonstrated the potential of the combination of two surveys in distinguishing between different cosmological models. We leave a detailed study of the synergies between these two surveys, as well as with other observational probes, in constraining cosmological parameters for a follow-up publication.

\begin{acknowledgments}
We are grateful to Ana Marta Pinho for helpful discussions on the subject of this work. This work was done in the context of project PTDC/FIS/111725/2009 (FCT, Portugal), with additional support from grant UID/FIS/04434/2013. CJM is supported by an FCT Research Professorship, contract reference IF/00064/2012, funded by FCT/MCTES (Portugal) and POPH/FSE (EC). MM is supported by the Foundation for Fundamental Research on Matter (FOM) and the Netherlands Organization for Scientific Research / Ministry of Science and Education (NWO/OCW). MM was also supported by the DFG TransRegio TRR33 grant on The Dark Universe during the preparation of this work. EC is supported by a STFC Rutherford Fellowship.

CJM and MM thank the Galileo Galilei Institute for Theoretical Physics for the hospitality and the INFN for partial support during the completion of this work.
\end{acknowledgments}

\bibliography{drift7}

\end{document}